\documentstyle[12pt]{article}

\def\op{\stackrel{\circ}{\otimes}}
\def\pp{+\!\!\!\!\!\smash{\supset}}
\newcounter{rown}[section]
\renewcommand{\theequation}{\thesection.\therown}
\begin{document}

\title{Quantum Deformation of the Poincare Supergroup and
$\kappa$-deformed Superspace }
\author{\em P. Kosi{\'n}ski \footnotemark[1]  \footnotemark[5]  ,
            J. Lukierski \footnotemark[2]  \footnotemark[6]  ,
            P. Ma{\'s}lanka \footnotemark[3]  \footnotemark[7]
         \    and J. Sobczyk \footnotemark[4]  \footnotemark[6]
}
\date{March 1994}
\maketitle
\thispagestyle{empty}

\begin{abstract}
The classical $r$-matrix for $N=1$ superPoincar{\'e} algebra, given by
Lukierski, Nowicki and Sobczyk is used to describe the graded Poisson
structure on the $N=1$ Poincar{\'e} supergroup. The standard
correspondence principle between the even (odd) Poisson brackets and
(anti)commutators leads to the consistent quantum deformation of the
superPoincar{\'e} group with the deformation parameter $q$ described by
fundamental
mass parameter $\kappa \quad (\kappa^{-1}=\ln{q})$. The $\kappa$-deformation
of $N=1$ superspace as
dual to the $\kappa$-deformed supersymmetry algebra is discussed.
\end{abstract}
\vskip 3cm
\footnoterule
{\small $^*$ Institute of Physics, University of
{\L}{\'o}d{\'z}, ul. Pomorska 149/153, 90-236 {\L}{\'o}d{\'z},
Poland\\
$^{\dag}$ Institute for Theoretical Physics, University of Wroc{\l}aw,
pl. Maxa Borna 9, 50-204 Wroc{\l}aw, Poland\\
$^{\ddag}$ Dept. of
Functional Analysis, Institute of Mathematics, University of {\L}{\'o}d{\'z},
ul. S. Banacha 22, 90-238 {\ \L}{\'o}d{\'z},
 Poland\\
$^{\S}$ Dipartamento di Fisica Teorica, University of Valencia, Av. dr.
Molimer 50, Burjasot, Valencia, Spain.\\
On leave of absence from Institute for Theoretical Physics, University
of Wroc{\l}aw, pl. Maxa Borna 9, 50-204 Wroc{\l}aw, Poland\\
$^{\P}$ Partially supported by {\L}{\'o}d{\'z} University grant 422/93\\
$^{\|}$ Partially
supported by KBN grant 2P30208706\\
$^{**}$ Partially supported by KBN grant 2P30221706p02}

\newpage \setcounter{page}{1}
\section{Introduction  }

\hskip .5cm
Recently in several papers
(\cite{Lukier91} - \cite{Chaichian}) there were
considered quantum deformations of $D=4$ Poincar{\'e} algebra which
describes the relativistic symmetries. Subsequently we would like to
stress here that
during the last twenty years the supersymmetric extensions of the
relativistic symmetries were one of the most studied ideas in the theory
of fundamental interactions. We conclude therefore that it is natural
to ask how do look
the quantum deformations of superalgebras or supergroups which describe
the supersymmetric extensions of the
four-dimensional space-time symmetries.

The deformation of $N=1$ superPoincar{\'e} algebra with fourteen
generators $I_A = ( M_i, L_i, P_{\mu}, Q_i, \overline {Q}_i ), \quad
(A=1, \ldots, 14)$ can be studied
at
least in two different ways:
\begin{description}
\item[a)] By considering the Hopf subalgebras of quantum
superconformal
algebra\linebreak $U_q(SU(2, 2; 1))$.

The complete description of this approach should take all possible
quantum deformations of $SU(2, 2; 1)$\footnote{We would like to recall
here that for the complexified conformal
algebra one can introduce the $R$-matrix with 7 parameters
\cite{Schir91}. The analogous general multiparameter deformations of
quantum superalgebras were not studied in the literature (see however
the partial results in \cite{Manin}).}. In the case studied so far (see
\cite{Dobrev92}) the minimal Hopf subalgebra of $U_q (SU(2,2;1))$
containing deformed $N=1$ superPoincar{\'e} generators has 16
generators; 14 generators of superPoincar{\'e} algebra ${\cal P}_{4;1}$
as well as the dillatation generator $D$ and the chiral generator $A$.
We have therefore
\begin{equation}
U_q (SU(2, 2; 1)) \supset U_q ({\cal P}_{4;1} \pp (D \oplus A))
\stepcounter{rown} \end{equation}
i. e. we obtain in such a way the quantum deformation of $D=1$
super-Weyl algebra.
\item[b)] By considering the contraction of quantum super-de Sitter
algebra ${\cal U}_q (OSp (1;4))$.

It appears that such a method provides a genuine 14-generator quantum
deformation of
$N=1$ Poincar{\'e} superalgebra, the $\kappa$-deformed
super-Poincar{\'e} algebra given firstly in \cite{LukSob93}, and
described briefly in Sect. 2.
\end{description}

In this paper we shall study further the quantum deformation of $N=1$
super-Poincar{\'e} group given in \cite{LukSob93}. From the $\kappa$-deformed
super-Poincar{\'e}
algebra, which is a non-commutative Hopf algebra, there can be extracted
the non-trivial classical $r$-matrix. Indeed, in \cite{LukSob93} it has
been shown that the graded-antisymmetric part of the coproducts in first
order in deformation parameter $h \equiv {1\over {\kappa}}$ is given by
\begin{equation}
\delta (X) = {1\over {\kappa}} [X \otimes {\bf 1} + {\bf 1} \otimes X,
r]
\stepcounter{rown} \end{equation}
\begin{equation}
r = L_i \wedge P_i - {i\over 4}Q_{\alpha} \wedge \overline {Q}_{\dot
{\alpha}} \equiv r^{AB} I_A \wedge I_B
\stepcounter{rown} \end{equation}
where $A \wedge B \equiv A \otimes B - (-1)^{\eta (A) \eta (B)}B \otimes
A; \quad i=1,2,3; \quad \alpha = 1,2$.\\
The bitensor $r \in \hat g \otimes \hat g$ given by (1.3) describes the
classical $r$-matrix for the $N=1$ Poincar{\'e} superalgebra, where
$L_i$ denotes the boost generators, $P_i$ - the three-momenta, and
$Q_{\alpha}, Q_{\dot {\alpha}}$ describe the supercharges written as
Weyl two-spinors. It appears that the classical $r$-matrix (1.3) satisfies the
graded {\it modified} classical Yang-Baxter equation $^{\small 2}$,
which permits to introduce consistently on
the space $g^*$ dual to $g$ the non-trivial multiplication structure,
determined by the cobracket (1.2). Introducing the generators $Z_A \in
\tilde g$ representing the supergroup parameters, one can define on the
functions $f(Z_A)$ the graded Poisson $r$-bracket$^{\small 2}$
\begin{equation}
\{f, g\} = \{f, g\}_R - \{f, g\}_L
\stepcounter{rown} \end{equation}
where $(a = R, L) ^{\small 3}$
\begin{equation}
\{f, g\}_a = (-1)^{\eta (A)\eta (B)}
(\stackrel{\leftarrow}{D}_A^{(a)}\!f)r^{AB}(\vec{D}_B^{(a)}\!g)
\stepcounter{rown} \end{equation}
and\\
\begin{description}
\item[-] $\stackrel{\leftarrow}{D}_A^{(a)}$ denotes left derivative
which is for $a=R \quad (a=L)$ right-invariant (left-invariant) under
supergroup transformations,
\item[-] $\vec{D}_A^{(a)}$ respectively denotes right derivative which
is
right-invariant (left-invariant) for $a=R (a=L)$.
\end{description}

In Sect.3 we shall consider more in detail the Poisson-Lie supergroup
structure on $N=1$ Poincar{\'e} supergroup. It appears that for the
choice of the $r$-matrix given by (1.3) the Poisson bracket (1.4) can be
consistently quantized in a standard way, by the substitution of
(graded) Poisson brackets by (anti-)commutators. In such a
way the supergroup parameters are promoted to the noncommuting
generators of
quantum $N=1$ Poincar{\'e} supergroup, with the
coproduct rules, described by the
composition law of two $N=1$ supersymmetry transformations.
\vskip .5cm
\footnoterule
{\noindent\small $^2$ For
non-supersymmetric case see \cite{Drin83} - \cite{Drin86}   \\
$^3$
In supersymmetric case one can introduce the
left- and right-side derivatives
$$
\vec{d}f = \vec{d}Z_A {{\vec{\partial}\!f}\over
{\partial\!Z_A}}\qquad \stackrel{\leftarrow}{d}\!f =
{{\stackrel{\leftarrow}{\partial}\!f}\over {\partial\!Z_A}}
\stackrel{\leftarrow}{d}\!a
\eqno (A.1)
$$
where ${\vec{d}}^2 = {\stackrel{\leftarrow}{d}}^2 = 0$, satisfying
different Leibnitz rules
$$
\vec{d}(f\!g) = \vec{d}\!fg +(-1)^{\eta(f)}f\vec{d}\!g\qquad
\stackrel{\leftarrow}{d}(f\!g) =
(-1)^{\eta(g)}\stackrel{\leftarrow}{d}\!fg +
f\stackrel{\leftarrow}{d}\!g
\eqno (A.2)
$$
One gets that
$$
{{\vec{\partial}\!f}\over {\partial\!Z_A}}=
(-1)^{\eta(f)\eta(Z_A)} {{\stackrel{\leftarrow}{\partial}\!f}\over
{\partial Z_A}}
\eqno (A.3)
$$
Using the relations (A.3) one can write the Poisson $r$-bracket  on a
supergroup in four different ways, which differ by suitable {\it sign}
factors. The choice (1.5) is the standard one.}
\pagebreak

It appears that after this quantization procedure the Lorentz sector of
the quantum $N=1$ Poincar{\'e} supergroup is classical - in analogy with
the case of quantum Poincar{\'e} group, considered previously by Zakrzewski
\cite{Zakrzinpress}. The deformation of the remaining generators of
quantum $N=1$ Poincar{\'e} supergroup, describing translations and
supertranslations, provides the $\kappa$-deformed $N=1$ superspace,
which is discussed in Sect.4. Finally in Sect.5 we present an outlook
and some unsolved problems.

\section{$D=4$ Quantum superPoincar{\'e} Algebra}

The $\kappa$-deformed $D=4$ Poincar{\'e} superalgebra given in
\cite{LukSob93} has the structure of noncommutative and noncocommutative
Hopf superalgebra. It is described by the following set of relations:
\begin{description}
\item[a)] Lorentz sector $(M_{\mu\nu} = (M_i, N_i)$, where $M_i =
\frac{1}{2}
\epsilon_{ijk}M_{jk}$ describe the non-relativistic $O(3)$ rotations,
and $N_i$ describe boosts).
  \begin{description}
\item[$i)$] {\it algebra}\\
\stepcounter{rown}
$$
[M_i, M_j] = i\epsilon_{ijk}M_k \qquad\quad  [M_i, L_j] = i\epsilon_{ijk}L_k
\eqno (\theequation a)
$$
$$
[L_i, L_j] = -i\epsilon_{ijk}(M_k \cosh{{P_0}\over {\kappa}} - {1\over
{8\kappa}}T_k \sinh{{P_0}\over {2\kappa}} + {1\over {16\kappa^2}}P_k
(T_0 - 4M))
\eqno (\theequation b)
$$
where  $(\mu = 0, 1, 2, 3)$
\begin{equation}
T_{\mu} = Q^A(\sigma_{\mu})_{A\dot B}Q^{\dot B}
\stepcounter{rown}
\end{equation}
\item[$ii)$] {\it coalgebra}\\
$$
\Delta (M_i) = M_i \otimes \hbox{\bf 1} + \hbox{\bf 1} \otimes M_i
\eqno (2.3 a) \stepcounter{rown}
$$
$$
\begin{array}{ll}
\Delta (L_i) = L_i \otimes e^{{P_0}\over {2\kappa}} + e^{-{{P_0}\over
{2\kappa}}}\otimes L_i + {1\over {2\kappa}}\epsilon_{ijk} (P_j
\otimes\null&\null\\ \\
\null\otimes M_k e^{{P_0}\over {2\kappa}} + M_j e^{-{{P_0}\over
{2\kappa}}}\otimes P_k)+\null&\null\\ \\
\null+{i\over {8\kappa}}
(\sigma_i)_{\dot\alpha\beta}(\overline{Q}_{\alpha}e^{-{{P_0}\over
{4\kappa}}}\otimes Q_{\beta}e^{{P_0}\over {4\kappa}} +
Q_{\beta}e^{-{{P_0}\over {4\kappa}}}\otimes
\overline{Q}_{\dot\alpha}e^{{P_0}\over {4\kappa}}&\null
\end{array}
\eqno (2.3 b)
$$
\item[$iii)$] {\it antipodes}\\
$$
\begin{array}{ll}
S(M_i)&= - M_i\\
S(N_i)&= - N_i + {{3i}\over {2\kappa}}P_i - {i\over
{8\kappa}}(Q\sigma_i\overline{Q} + \overline{Q}\sigma_i Q)
\end{array}
\eqno (2.4)
$$
\stepcounter{rown}
  \end{description}
\item[b)] Fourmomenta sector $P_{\mu} = (P_i, P_0)$
  \begin{description}
\item[$i)$] {\it algebra}\\
$$
\lbrack M_i, P_j\rbrack  = i\epsilon_{ijk}P_k  \qquad \lbrack M_j, P_0\rbrack
=
0
\eqno (2.5 a)
$$
\stepcounter{rown}\vskip -0.3cm
$$
\lbrack N_i, P_j\rbrack   = i\kappa\delta_{ij}\sinh{{P_0}\over {\kappa}}
\qquad
\lbrack N_i, P_0\rbrack   = iP_i
\eqno (2.5 b)
$$
$$
\lbrack P_{\mu}, P_{\nu}\rbrack  = 0  (\mu , \nu = 0, 1, 2, 3)
\eqno (2.5 c)
$$
\item[$ii)$] {\it coalgebra}\\
$$
\Delta (P_i)  = P_i \otimes e^{{P_0}\over {2\kappa}} + e^{-{{P_0}\over
{2\kappa}}} \otimes P_i
\eqno (2.6 a)
$$
$$
\Delta (P_0)  = P_0 \otimes \hbox{\bf 1} + \hbox{\bf 1} \otimes P_0
\eqno (2.6 b)
$$
  \end{description}
The antipode is given by the relation
\stepcounter{rown}
$S(P_{\mu}) = -P_{\mu}$.
\item[c)] Supercharges sector \cite{LukSob93}
  \begin{description}
\item[$i)$] {\it algebra}
$$
\begin{array}{ll}
\{Q_{\alpha}, Q_{\dot\beta}\} & =
4\kappa\delta_{\alpha\beta}\sin{{P_0}\over {2\kappa}} - 2P_i
(\sigma_i)_{\alpha\dot\beta} \\ \\
\{Q_{\alpha}, Q_{\beta}\} & = \{Q_{\dot\alpha}, Q_{\dot\beta}\} = 0
\end{array}
\eqno(2.7 a)
$$\stepcounter{rown}
$$
\lbrack M_i, Q_{\alpha}\rbrack = -{1\over 2} (\sigma _i)_{\alpha}^{\dot\beta}
Q_{\beta} \quad
\lbrack M_i, Q_{\dot\alpha}\rbrack = -{1\over 2} (\sigma
_i)_{\dot\alpha}^{\dot\beta} Q_{\dot\beta}
\eqno (2.7 b)
$$
$$
\lbrack N_i, Q_{\alpha}\rbrack = -{i\over 2}\cosh{{P_0}\over {2\kappa}} (\sigma
_i)_{\alpha}^{\beta}Q_{\beta}\quad
\lbrack N_i, Q_{\dot\alpha}\rbrack = {i\over 2}\cosh{{P_0}\over {2\kappa}}
(\sigma
_i)_{\dot\alpha}^{\dot\beta}Q_{\dot\beta}
\eqno (2.7 c)
$$
$$
\begin{array}{ll}
\lbrack P_{\mu}, Q_{\alpha}\rbrack  & = \lbrack P_{\mu}, Q_{\dot\beta}\rbrack
=
0
\end{array}
\eqno (2.7 d)
$$
\item[$ii)$] {\it coalgebra}
\begin{equation}
\begin{array}{ll}
\Delta (Q_{\alpha}) & = Q_{\alpha} \otimes e^{{P_0}\over {4\kappa}} +
e^{-{{p_0}\over {4\kappa}}} \otimes Q_{\alpha}\\
\Delta (Q_{\dot\alpha} & = Q_{\dot\alpha} \otimes e^{{P_0}\over {4\kappa}} +
e^{-{{P_0}\over {4\kappa}}} \otimes Q_{\dot\alpha}
\end{array}
\stepcounter{rown} \end{equation}
\item[$iii)$] {\it antipodes}
\begin{equation}
S(Q_{\alpha}) = - Q_{\alpha} \qquad S(Q_{\dot\alpha}) = - Q_{\dot\alpha}
\stepcounter{rown} \end{equation}
  \end{description}
\end{description}
On the basis of the relations (2.3) - (2.7) one can
single out the following features of the quantum superalgebra
${\cal U}_{\kappa} ({\cal P}_{4;1})$ :
\begin{description}
\item[$i)$] The algebra coproducts and antipodes of Lorentz boosts $N_i$
do depend on $Q_{\alpha}, Q_{\dot\alpha}$ i.e. the $\kappa$-deformed
Poincar{\'e} as well as Lorentz sectors do not form the Hopf
subalgebras.
\item[$ii)$] Putting in the formulae (2.1) - (2.6) $Q_{\alpha} =
Q_{\dot\alpha} = 0$ one obtains the $\kappa$-deformed Poincar{\'e}
algebra considered in \cite{Lukier92}, i.e.
$$
{\cal U}_{\kappa} ({\cal P}_{4;1}) \left |_{Q_{\alpha} = Q_{\dot\alpha} =
0}\right . = {\cal U}_{\kappa}({\cal P}_4)
$$
\item[$iii)$] From (2.5 c) we see that the fourmomenta commute. This
property implies by duality the standard addition formula for the
space-time fourvectors (see Sect.4).
\end{description}

\section{Poisson $r$-brackets For $N=1$ Poincar{\'e} Supergroup And
Their Quantization}

The classical $N=1$ Poincar{\'e} Lie
superalgebra with the cobracket (1.2) describes the $N=1$ Poincar{\'e}
Lie super-bialgebra $(\hat g, \hat\delta)$, which is called {\it coboundary}
\cite{Drin86} due to the relation (1.3) between the cobracket $\delta$
and the $r$-matrix.

The coboundary super-bialgebras with the $r$-matrix satisfying the
modified classical Yang-Baxter equation describe infinitesimally
Poisson-Lie supergroups, with the supergroup action $(Z_A, Z_B)
\longrightarrow Z_A \circ Z_B$ consistent with the Poisson structure
given by the $r$-Poisson bracket (1.5). These brackets satisfy the
following properties:
\begin{enumerate}
\item Graded antisymmetry
\begin{equation}
\{f, g\} = - (-1)^{\eta\!(f)\eta\!(g)} \{g, f\}
\stepcounter{rown} \end{equation}
\item Graded Jacobi identity
\begin{eqnarray}
\lefteqn{(-1)^{\eta\!(f)\eta\!(h)} \{f, \{g, h\}\} +
(-1)^{\eta\!(g)\eta\!(h)}\cdot\null}\nonumber\\
&&\null\cdot\{h, \{f, g\}\} + (-1)^{\eta\!(f)\eta\!(g)} \{g, \{h, f\}\} =
0                                   \stepcounter{rown}
\end{eqnarray}
\item Graded Leibnitz rules
\begin{equation}
\begin{array}{ll}
\{f, gh\} & = \{f, g\} h + (-1)^{\eta\!(f)\eta\!(g)}g\{f, h\} \\
\\
\{fg, h\} & = f\{ g, h\} + (-1)^{\eta\!(g)\eta\!(h)}\{ f, h\} g
\end{array}
\stepcounter{rown} \end{equation}
\item Lie -- Poisson property

Let us write the coproduct induced by the composition law of two supergroup
transformations
\begin{equation}
\Delta (Z) = Z\op Z
\stepcounter{rown} \end{equation}
where "$\op$" denotes that we take the composition rule described by
"$\circ$" and replace the product by the tensor product. The Lie-Poisson
property takes the form
\begin{equation}
\Delta \{f, g\} = \{\Delta (f), \Delta (g)\}
\stepcounter{rown} \end{equation}
where the following rule for the multiplication of graded tensor
products should be used:
\begin{equation}
(f_1 \otimes f_2) (g_1 \otimes g_2) =
(-1)^{\eta\!(f_2)}(-1)^{\eta\!(g_1)}f_1g_1 \otimes f_2g_2
\stepcounter{rown} \end{equation}
\end{enumerate}

In order to calculate explicitly the Poisson bracket (1.4)
one can express the right- and left-invariant derivatives in terms of the
ordinary ones,
i.e. rewrite (1.4) as follows
\begin{equation}
\{f, g\} = f {{\stackrel{\leftarrow}{\partial}}\over {\partial\!Z_A}}
\omega^{AB}(z) {{\vec{\partial}}\over {\partial\!Z_B}}g
\stepcounter{rown} \end{equation}
If we observe that right
\begin{equation}
\stackrel{\leftarrow}{D}_A^{(a)} =
{{\stackrel{\leftarrow}{\partial}}\over {\partial\!Z^B}}
\stackrel{\leftarrow}{\mu}^{(a)B}_A (Z) \qquad
\vec{D}_A^{(a)} =
{\vec{\mu}^{(a)B}_A}(Z){{\vec{\partial}}\over {\partial\!Z^B}}
\stepcounter{rown} \end{equation}
where $\stackrel{\leftarrow}{\mu}^{(a)}, {\vec{\mu}}^{(a)}$
can be calculated by the differentiation of the composition formulae of
the supergroup parameters $Z_A$,
one obtains that $(L=+,\quad R=-)$:
\begin{equation}
\omega^{AB}(Z) = \stackrel{\leftarrow}{\mu}^{(+)A}_C(Z)
r^{CD}{\vec{\mu}^{(+)B}_D}(Z) - \stackrel{\leftarrow}{\mu}^{(-)A}_C (Z)
r^{CD}\vec{\mu}_D^{(+)B}(Z)
\stepcounter{rown} \end{equation}
where the leading term at $Z=0$ is linear , and describes the cobracket
of the $N=1$ Poincar{\'e} bi-superalgebra $(\hat g, \hat\delta)$, in
accordance with the relation (1.2).

The quantization of the $N=1$ superPoincar{\'e} algebra consists in
two steps:
\hskip -0.3cm
\begin{enumerate}
\item Write (3.9) for the independent parameters $Z^A$ (the generators
of the algebra of functions on the supergroup ${\cal P}_{4;1}$)
\begin{equation}
\{Z^A, Z^B\} = \omega^{AB}(Z)
\stepcounter{rown} \end{equation}
and calculate $\omega^{AB}$ by choosing the functions
$\stackrel{\leftarrow}{\mu}^{(a)}, \vec{\mu}^{(a)}$ in
(3.8), depending on the parametrization of the supergroup.
\item Quantize the Poisson bracket by the substitution
\begin{equation}
\{Z^A, Z^B\} \longrightarrow \left\{ \begin{array}{ll}
{1\over {i\hbar}}[{\hat Z}^A, {\hat Z}^B]_- & \hbox{if} \quad
\eta\!(A)\cdot\eta\!(B) = 0\\
{1\over {i\hbar}}[{\hat Z}^A, {\hat Z}^B]_+ & \hbox{if} \quad
\eta\!(A)\cdot\eta\!(B) = 1
\end{array}\right.
\stepcounter{rown} \end{equation}
where $[{\hat A}, {\hat B}]_{\pm} = {\hat A}{\hat B} \pm {\hat B}{\hat
A}$, and choose the ordering of the ${\hat Z}$-variables in
$\omega^{AB}$ in such a way that the Jacobi identities are satisfied,
and the coproduct (3.4) is a homomorphism of
the quantized superalgebra.
\end{enumerate}

Let us recall
the supergroup composition law ($A$ is $2\times 2\quad Sl (2;{\bf C})$
matrix).
\begin{eqnarray} \lefteqn{(X_{\mu}, \theta_{\alpha}, A_{\alpha}^{\beta})
\circ (X_{\mu}', \theta_{\alpha}', A_{\alpha}'^{\beta})=\null}\nonumber\\
&&\null=(X_{\mu}+\Lambda_{\mu}^{\nu}(A)X_{\nu}'+{i\over
2} (\theta'^T A^{-1}\sigma^{\mu}\overline\theta -
\theta^T\sigma^{\mu}(A^+)^{-1}\overline\theta'),\nonumber\\
&&\theta_{\alpha} +\theta_{\beta}'(A^{-1})^{\beta}_{\alpha}, \quad
A_{\alpha}^{\gamma}{A'}_{\gamma}^{\beta})
\stepcounter{rown}
\end{eqnarray}
The formulae (3.12) permits to calculate the functions
$\stackrel{\leftarrow}{\mu}^{(\pm)}, \vec{\mu}^{(\pm)}$ in the formula
(3.9). We obtain for example the following formulae for left-sided
left-invariant super-derivatives:
\begin{equation}
\begin{array}{lll}
\vec{D}_{\alpha}^{(+)}&=&(A^{-1})_{\alpha}^{\beta} {{\partial}\over
{\partial\theta^{\beta}}}+{i\over
2}(A^{-1}\sigma^{\mu}\overline\theta_{\alpha}) {{\partial}\over
{\partial\!X^{\mu}}}\\
{\vec{D}^{(+)\beta}_{\alpha}}&=&A_{\gamma}^{\beta}{{\partial}\over
{\partial\!A_{\gamma}^{\alpha}}}
\end{array}
\stepcounter{rown} \end{equation}
and by conjugation
\begin{equation}
\begin{array}{lll}
\vec{D}_{\dot\alpha}^{(+)}&=&(A^{-1})_{\dot\alpha}^{\dot\beta}{{\partial}\over
{\partial\overline\theta ^{\dot\beta}}} + {i\over 2}(\theta^T \sigma^{\mu}
(A^+)^{-1})_{\dot\alpha} {{\partial}\over {\partial\!X^{\mu}}}\\
{\vec{D}^{(+)\dot\beta}_{\dot\alpha}}&=&(A)_{\dot\gamma}^{\dot\beta}
{{\partial}\over {\partial\!A_{\dot\gamma}^{\dot\alpha}}}
\end{array}
\stepcounter{rown} \end{equation}
Calculating the remaining invariant derivatives on the bosonic
Poincar{\'e} subgroup and
inserting in the formula (3.9) the $r$-matrix (1.3)
we obtain the following fundamental $r$-Poisson brackets for the
coordinates $(X_{\mu}, A_{\alpha}^{\beta}, A_{\dot\alpha}^{\dot\beta},
\theta_{\alpha}, \theta_{\dot\alpha})$ on $N=1$ Poincar{\'e}
supergroup\setcounter{footnote}{3}
\footnote{We use the spinorial representation of the Lorentz generators,
e.g.
$
L_i = {1\over
4}(\sigma_i)_{\alpha}^{\beta}L^{\alpha}_{\beta}+(\overline\sigma_i)_{\dot\alpha}
^{\dot\beta}
L^{\dot\alpha}_{\dot\beta}
$}:
\begin{description}
\item[a)] Lorentz sector $(A_{\alpha}^{\beta},
A_{\dot\alpha}^{\dot\beta})$\\
The Lorentz subgroup parameters are classical, i.e.
\begin{equation}
\{A_{\alpha}^{\beta}, A_{\gamma}^{\delta}\} = \{A_{\alpha}^{\beta},
A_{\dot\gamma}^{\dot\delta} \} = \{A_{\dot\alpha}^{\dot\beta},
A_{\dot\gamma}^{\dot\beta} \} = 0
\stepcounter{rown} \end{equation}
\item[b)] Translations $(X_{\mu})$ (we denote $\theta =
{{\theta_1}\choose{\theta_2}}, \overline\theta = {{\theta_{\dot
1}}\choose{\theta_{\dot 2}}}$)
\begin{equation}
\begin{array}{ll}
\{X^i, X^j\} = {i\over {8\kappa}}\theta^T
\sigma^i (\hbox{\bf 1}_2 - (AA^+)^{-1})\sigma^j
\overline\theta-{i\over {8\kappa}}\theta^T \sigma^j (\hbox{\bf
1}-(AA^+)^{-1})\sigma^i \overline\theta&\null\\ \\
\{X^0, X^j\} = -{i\over {\kappa}}X^j + {i\over
{8\kappa}}\theta^T \lbrack \sigma^j,
(AA^+)^{-1}\rbrack \overline\theta&\null
\end{array}
\stepcounter{rown}
\end{equation}
\begin{equation}
\begin{array}{lll}
\{ A_{\alpha}^{\beta}, X^i\}&=&{1\over
{2\kappa}}((A\sigma_{n})_{\alpha}^{\beta} \Lambda^i_n (A) -
(\sigma^i\cdot\!A)_{\alpha}^{\beta})\\            \\
\{ A_{\alpha}^{\beta}, X^0\}&=&{1\over {2\kappa}}(A\sigma_i
)_{\alpha}^{\beta} \Lambda^0_i (A)
\end{array}
\stepcounter{rown} \end{equation}
\item[c)] Supertranslations
\begin{equation}
\{ \theta^{\alpha}, \theta^{\beta}\}=\{ \theta^{\dot\alpha},
\theta^{\dot\beta}\} = 0\quad
\{ \theta^{\alpha}, \theta^{\dot\beta}\}={i\over {2\kappa}}(\hbox{\bf
1} - AA^+)^{-1})^{\dot\beta \alpha}
\stepcounter{rown} \end{equation}
\begin{equation}
\begin{array}{lll}
\{ X^i, \theta_{\alpha} \} &=&{1\over
{4\kappa}}(\theta^T\sigma^i)_{\gamma} (\hbox{\bf 1}_2 -
(AA^+)^{-1})^{\gamma}_{\alpha}\\ \\
\{ X^0, \theta_{\alpha}\}&=&-{1\over
{4\kappa}}\theta^T_{\gamma}(\hbox{\bf 1}_2 +
(AA^+)^{-1})^{\gamma}_{\alpha}
\end{array}
\stepcounter{rown} \end{equation}

\begin{equation}
\begin{array}{lll}
\{ A_{\alpha}^{\beta}, \theta^{\gamma}\}&=&\{
A_{\dot\alpha}^{\dot\beta}, \theta^{\gamma} \} = 0
\end{array}
\stepcounter{rown} \end{equation}
\end{description}

In order to quantize the Poisson brackets (3.15 - 3.20) we perform the
substitution (3.11). It appears that this substitution is consistent
with Jacobi identities if we keep the order of the coordinate generators
on {\it rhs} of (3.16) also in quantized case\nolinebreak\footnote{For other
relations (3.17-3.20) the problem does not occur due to the classical
nature of the Lorentz sector (see (3.16)).}. Furthermore, rewriting the
composition
law (3.12) as the coproduct rule for the coordinate generators, i.e.
\begin{equation}
\begin{array}{ll}
\Delta (X_{\mu})&=X_{\mu}\otimes\hbox{\bf 1} +
\Lambda_{\mu}^{\nu}(A)\otimes X_{\nu}-
{i\over 2}({A^{-1}_{\alpha}}^{\beta} \sigma^{\mu}_{\beta\dot\gamma}
\theta^{\dot\gamma} \otimes \theta^{\alpha} +
\theta^{\alpha}\sigma^{\mu}_{\alpha\dot\beta}A^{-1\beta}_{\dot\gamma}
\otimes \theta^{\dot\gamma})\\
\Delta (\theta_{\alpha})&=\theta_{\alpha}\otimes\hbox{\bf
1}+(A^{-1})^{\beta}_{\alpha}\otimes\theta_{\beta}\\
\Delta (A_{\alpha}^{\beta})&=A_{\alpha}^{\gamma}\otimes
A_{\gamma}^{\beta}
\end{array}
\stepcounter{rown}
\end{equation}
One can show that the formulae (3.21) describe the homomorphism of the
quantized
superalgebra given in Sect.2. Adding the formulae for the antipodes
\begin{equation}
S(X^{\mu})=-\Lambda^{\mu}_{\nu}(A^{-1})X^{\nu}\quad
S(A_{\alpha}^{\beta})=(A^{-1})_{\alpha}^{\beta}
\stepcounter{rown} \end{equation}
$$
S(\theta^{\alpha})=-A_{\beta}^{\gamma}\theta^{\beta}
$$
we see that
we have obtained the complete set of relations describing
the $\kappa$-deformation of $N=1$ Poincar{\'e} supergroup.

Let us observe that
\begin{description}
\item[a)] If we put $A^+A=1$, i.e. we consider the
semidirect product $T_{4;4} \pp SU(2)$ of the quantum subgroup
$T_{4;4}$ (quantum fourtranslations + quantum supertranslations) and
$SU(2)$ describing the space
rotations, only in two relations (first relation (3.17) and second
relation (3.19)) the nontrivial $\kappa$-deformation occurs.
\item[b)] If we put $A=1$, i.e. we consider the
quantum subgroup $T_{4;4}$, we obtain the $\kappa$-deformed
$N=1$ superspace. It appears that only the commutator $[X^0,
\theta_{\alpha}]$ is $\kappa$-deformed.
\item[c)] Putting in (3.15) - (3.17)
$\theta^{\alpha} = \theta^{\dot\alpha} = 0$ one
recovers the $\kappa$-deformed inhomogeneous $ISl(2;\hbox{\bf C})$ group,
given in
\cite{Masl93}.
\end{description}

\section{$\kappa$-deformed $N=1$ Superspace}

Let us recall firstly that for $\kappa$-deformed relativistic theory,
with infinitesimal symmetries described by the $\kappa$-deformed
Poincar{\'e} algebra \cite{Lukier91,Giller92,Lukier92,Lukier93,Bacry93}
there are two different ways of introducing the Poincar{\'e} group and
space-time coordinates:
\begin{description}
\item[a)]
Using the formula (2.5 c) one can consider the space-time coordinates by
considering {\it ordinary} Fourier transforms of the functions depending
on the commuting fourmomenta \cite{Lukier92,Lukier93,Giller93}. In such
an approach the space-time coordinate operators $\hat X_{\mu}$ commute
and are introduced as the operators satisfying the relations
\begin{equation}
[\hat X_{\mu}, \hat P_{\nu}]=i\eta_{\mu\nu}
\stepcounter{rown} \end{equation}
\item[b)]  Using the duality relation for Hopf algebras described by
the scalar product on quantum double with the following properties
\begin{equation}
\begin{array}{lll}
<\Delta (\hat z)|\hat g_1 \otimes \hat g_2>&=&<\hat z |\hat g_1 \hat
g_2>\\    \\
<\hat z_1 \otimes \hat z_2|\Delta (\hat g)>&=&<\hat z_1 \hat z_2 |\hat
g>
\end{array}
\stepcounter{rown} \end{equation}
we easily see that for standard duality relation between $\hat X_{\mu}$
and $\hat P_{\mu}$ generators
\begin{description}
\item[-]  non-cocommutative fourmomenta (see (2.6)) imply the
non-commutativity of the coordinates \cite{Zakrzinpress}:
\begin{equation}
\lbrack \hat X^i, \hat X^j\rbrack = 0\qquad
\lbrack \hat X^0, \hat X^j\rbrack = {1\over {\kappa}}\hat X^j
\stepcounter{rown} \end{equation}
\item[-]  commutativity of the fourmomenta imply that
\begin{equation}
\Delta (\hat X^{\mu}) = \hat X^{\mu} \oplus \hbox{\bf 1} + \hbox{\bf 1}
\oplus \hat X^{\mu}
\stepcounter{rown} \end{equation}
\end{description}
One can rewrite the coproduct formulae (2.6) and (4.4) as the addition
formulae for the fourmomenta
$$
p_i^{(1+2)}=p_i^{(1)}e^{{p_0^{(2)}}\over {2\kappa}} +
p_i^{(2)}e^{-{{p_0^{(1)}}\over {2\kappa}}} \qquad
p_0^{(1+2)}=p_0^{(1)} + p_0^{(2)}
    \stepcounter{rown}
\eqno (\theequation a)
$$
and for the space-time coordinates
$$
x^{\mu}_{(1+2)}=x^{\mu}_{(1)}+x^{\mu}_{(2)}
\eqno (\theequation b)
$$
If we introduce the following element of the quantum double describing
the translation sector of $\kappa$-Poincar{\'e} $(\hat X^0 = -\hat
X_0,\quad  \hat X^i = -\hat X_i)$\footnote{For the concepts of
exponentiation of the generators of quantum double, consisting of
quantum Lie algebra and dual quantum Lie group see
\cite{Fronsdal93} - \cite{Bonechi93}, where the exponentials (4.6) are
called quantum $T$-matrices. The notion of quantum $T$-matrix
is related to
the notion of the universal bicharacter of Woronowicz
(see e.g.\cite{Woron91}).}
\begin{equation}
G(\hat X^{\mu}; \hat P_{\mu})=
e^{-\frac{i}{2}\hat X_0 \otimes \hat P_0} e^{i\hat X_i \otimes \hat P_i}
e^{-\frac{i}{2}\hat X_0 \otimes \hat P_0}
\stepcounter{rown} \end{equation}
one can encode the additional formulae (4.5a-b) into the following
multiplication rules
$$
G(\hat X^{\mu}; p_{\mu}^{(1)})G(\hat X^{\mu}; p_{\mu}^{(2)})\quad =\quad G(\hat
X^{\mu}; p_{\mu}^{(1+2)})
\stepcounter{rown}
\eqno (\theequation a)
$$
$$
G(x^{\mu}_{(1)}; \hat P_{\mu})G(x^{\mu}_{(2)}; \hat P_{\mu})\quad =\quad G
(x^{\mu}_{(1+2)}; \hat P_{\mu})
\eqno (\theequation b)
$$
We see therefore that the relations (4.6) describe the generalization of
Fourier transform kernels to the case of the translation sector of
$\kappa$-Poincar{\'e} group,
with the coproducts determining their multiplication rule.
\end{description}

Let us extend such a scheme to $N=1$ superPoincar{\'e} case. The
non-commutative Hopf algebra, describing $\kappa$-deformed superspace,
is obtained by the quantization of the relations (3.16)-(3.19) with
$A=1$. One
obtains
$$
\lbrack \hat X^i, \hat X^j\rbrack = 0\qquad
\lbrack \hat X^0, \hat X^j\rbrack = {1\over {\kappa}}\hat X^j
$$
\begin{equation}
\begin{array}{lll}
\{\hat\theta^{\alpha}, \hat\theta^{\beta}\}&=&\{\hat\theta^{\alpha},
\hat\theta^{\dot\beta}\}=\{\hat\theta^{\dot\alpha},
\hat\theta^{\dot\beta} \} = 0\\ \\
\lbrack \hat X^i, \hat\theta^{\alpha}\rbrack &=&\lbrack \hat X^i,
\hat\theta^{\dot\alpha}\rbrack =0
\end{array}
\stepcounter{rown} \end{equation}
$$
\lbrack \hat X^0, \hat\theta^{\alpha}\rbrack = -{1\over
{2\kappa}}\hat\theta^{\alpha}\qquad
\lbrack \hat X^0, \hat\theta^{\dot\alpha}\rbrack = -{1\over
{2\kappa}}\hat\theta^{\dot\alpha}
$$
and the coproducts (3.26) implying the following composition law in
superspace:
\begin{equation}
\begin{array}{ll}
\hat\theta^{\alpha}_{(1+2)}&=\hat\theta^{\alpha}_{(1)}+\hat\theta^{\alpha}_{(2)}
\qquad
\hat\theta^{\dot\alpha}_{(1+2)}=\hat\theta^{\dot\alpha}_{(1)}+\hat\theta^{\dot\a
lpha}_{(2)}\\
\\
\hat X^{\mu}_{(1+2)}&=X^{\mu}_{(1)}+X^{\mu}_{(2)}+{i\over
2}(\sigma^{\mu})_{\alpha\dot\beta}
(\theta^{\dot\beta}_{(1)}\theta^{\alpha}_{(2)}-\theta^{\alpha}_{(1)}\theta^{\dot
\beta}_{(2)})
\end{array}
\stepcounter{rown} \end{equation}
We recall that $\kappa$-deformed $N=1$ superalgebra is described by the
relations (2.7) and the coproducts (2.8). The addition formula of the
Grassmann-algebra-valued eigenvalues $q_{\alpha}, q_{\dot\alpha}$ of the
supercharges, induced by (2.12), is the following
\begin{equation}
\begin{array}{lll}
q_{\alpha}^{(1+2)}&=&q_{\alpha}^{(1)}e^{{p_0^{(2)}}\over
{4\kappa}}+q_{\alpha}^{(2)}e^{-{{p_0^{(1)}}\over {4\kappa}}}\\
q_{\dot\alpha}^{(1+2)}&=&q_{\dot\alpha}^{(1)}e^{{p_0^{(2)}}\over
{4\kappa}}+q_{\dot\alpha}^{(2)}e^{-{{p_0^{(1)}}\over {4\kappa}}}
\end{array}
\stepcounter{rown} \end{equation}
If we introduce the following quantum counterpart of the finite
supertranslation
group elements in momentum as well as coordinate superspace
$$                    \stepcounter{rown}
G(p_{\mu}, q_{\alpha}, q_{\dot\alpha})=e^{-\frac{i}{2}\hat X_0 p_0}e^{i(\hat
X^{i}p_{i}+\hat\theta^{\alpha}q_{\alpha}+\hat\theta^{\dot\alpha}q_{\dot\alpha})}
e^{-\frac{i}{2}\hat X_0 p_0}
\eqno (\theequation a)
$$
$$
\tilde G (x_{\mu}, \theta_{\alpha},
\theta_{\dot\alpha})=e^{i(x^{\mu}{\tilde P}_{\mu} +
\theta^{\alpha}Q_{\alpha}+\theta^{\dot\alpha}Q_{\dot\alpha})}
\eqno (\theequation b)\stepcounter{rown}
$$
where $\tilde P_0 = 2\kappa \sinh{\frac{P_0}{2\kappa}}$ and $\tilde P_i
= P_i$,
we obtain the following multiplication laws:
$$
G(p_{\mu}^{(1)}, q_{\alpha}^{(1)}, q_{\dot\alpha}^{(1)})
G(p_{\mu}^{(2)}, q_{\alpha}^{(2)}, q_{\dot\alpha}^{(2)})   =
G(p_{\mu}^{(1+2)}, q_{\alpha}^{(1+2)}, q_{\dot\alpha}^{(1+2)})
\eqno (\theequation a)
$$
$$
\tilde G(x^{\mu}_{(1)}, \theta^{\alpha}_{(1)}, \theta^{\dot\alpha}_{(1)})
\tilde G(x^{\mu}_{(2)}, \theta^{\alpha}_{(2)},
\theta^{\dot\alpha}_{(2)}) =
\tilde G(x^{\mu}_{(1+2)}, \theta^{\alpha}_{(1+2)},
\theta^{\dot\alpha}_{(1+2)})
\eqno (\theequation b)
$$
Following the discussion for ordinary supersymmetry (see e.g.
\cite{Ferrara74}) one can consider the objects (4.11 a) and (4.11 b) as
describing respectively the superfields in momentum superspace and in
the usual (coordinate) superspace.

It should be mentioned that the algebra (4.8) describes the superspace
coordinates in the particular Lorentz frame $(A=1)$. If we allow
nontrivial Lorentz transformations, the algebra of superspace
coordinates is no longer closed, and one should consider the full
algebra given by (3.15-20).

\section{Outlook}
In this paper we presented quantum $\kappa$-deformation of $N=1$
Poincar{\'e} supergroup, which is a non-commutative and
non-co-commutative Hopf superalgebra. We would like to mention the
following problems which should be further studied:
\begin{description}
\item[i)]  It appears that for the non-semisimple Lie
(super)algebras the "naive" quantization (see (3.11))of the $r$-Poisson bracket
may be very
useful as a consistent quantization scheme.
In \cite{Zakrzinpress} as well as in the case presented in this paper
the ambiguities related to the ordering of the $rhs$ of the quantized
$r$-Poisson brackets are resolved in the unique way.
It is interesting to
classify for non-semisimple Lie (super)algebras the classical
$r$-matrices and find for which cases the "naive" quantization of the
$r$-Poisson bracket leads to a consistent quantization\footnote{This
programme is now under consideration, where also the classical
$r$-matrices for simple quantum Lie-algebras and the "naive"
quantization of corresponding quadratic $r$-Poisson brackets are
studied.}.
\item[ii)]  One can show that the $\kappa$-deformed $N=1$
supersymmetry algebra $(Q_{\dot\alpha}, Q_{\alpha}, P_{\mu})$ as a Hopf
superalgebra (see Sect.2) is dual to the Hopf superalgebra describing the $N=1
\quad \kappa$-deformed superspace (see Sect.4). It would be important to show
that the
whole $N=1 \quad \kappa$-deformed supergroup is dual (possibly modulo some
nonlinear
transformations of the generators) to the $N=1\quad \kappa$-Poincar{\'e}
superalgebra, given in \cite{LukSob93}. We would like to stress that
such duality for $D=4\quad\kappa$-deformed Poincar{\'e} group given in
\cite{Zakrzinpress} is not known.
\item[iii)]  It would be interesting to generalize the results of
\cite{LukSob93} and of this paper to $N>1$. We would like to mention
that complete $N$-extended Poincar{\'e} superalgebra, with $N(N-1)$
central charges, can be obtained by the construction of the superalgebra
$OSp(2N;4)$ \cite{Lukier82}. Replacing the classical superalgebra
$OSp(2N;4)$ by its $q$-analogue ${\cal U}_q(OSp(2N;4))$ and performing
quantum de-Sitter construction limit with the rescalling (2.3) one
should obtain the quantum deformation of $N$-extended superPoincar{\'e}
algebra. For obtaining $N$-extended $\kappa$-deformed Poincar{\'e}
supergroup it is sufficient to extend the classical $r$-matrix (1.3) to
$N>1$ and follow the method presented in this paper.
\end{description}


\end{document}